\documentclass[sigconf]{acmart}
\renewcommand\footnotetextcopyrightpermission[1]{} 
\AtBeginDocument{%
  \providecommand\BibTeX{{%
    \normalfont B\kern-0.5em{\scshape i\kern-0.25em b}\kern-0.8em\TeX}}}

\begin{document}


\title{GPT-4 and Safety Case Generation: An Exploratory Analysis}
\author{Mithila Sivakumar}
\affiliation{%
  \institution{York University}
  \streetaddress{4700 Keele St}
  \city{Toronto}
  \country{Canada}}
\email{msivakum@yorku.ca}

\author{Alvine Boaye Belle}
\affiliation{%
  \institution{York University}
  \streetaddress{4700 Keele St}
  \city{Toronto}
  \country{Canada}}
\email{alvine.belle@lassonde.yorku.ca}

\author{Jinjun Shan}
\affiliation{%
  \institution{York University}
  \streetaddress{4700 Keele St}
  \city{Toronto}
  \country{Canada}}
\email{jjshan@yorku.ca}

\author{Kimya Khakzad Shahandashti}
\affiliation{%
  \institution{York University}
  \streetaddress{4700 Keele St}
  \city{Toronto}
  \country{Canada}}
\email{kimya@yorku.ca}




\begin{abstract}
In the ever-evolving landscape of software engineering, the emergence of large language models (LLMs) and conversational interfaces, exemplified by ChatGPT, is nothing short of revolutionary. While their potential is undeniable across various domains, this paper sets out on a captivating expedition to investigate their uncharted territory—the exploration of generating safety cases. 
In this paper, our primary objective is to delve into the existing knowledge base of GPT-4, focusing specifically on its understanding of the Goal Structuring Notation (GSN), a well-established notation allowing to visually represent safety cases. Subsequently, we perform four distinct experiments with GPT-4. These experiments are designed to assess its capacity for generating safety cases within a defined system and application domain. To measure the performance of GPT-4 in this context, we compare the results it generates with ground-truth safety cases created for an X-ray system system and a Machine-Learning (ML)-enabled component for tire noise recognition (TNR) in a vehicle. This allowed us to gain valuable insights into the model's generative capabilities. Our findings indicate that GPT-4 demonstrates the capacity to produce safety arguments that are moderately accurate and reasonable. Furthermore, it exhibits the capability to generate safety cases that closely align with the semantic content of the reference safety cases used as ground-truths in our experiments.

\end{abstract}



\maketitle
\pagestyle{empty} 

\section{Introduction}

Safety incidents involving autonomous vehicles and other safety-critical systems can lead to fatal outcomes and pose significant risks to both manufacturers and the public \cite{b13}. To address this, safety-critical systems, including cyber-physical systems, are regulated through safety standards like ISO 26262, IEC 61508, and DO-178C \cite{b8, b14, b32}. Such standards usually require the construction of a \textbf{\textit{"safety case"}} to support the certification of these systems. This allows verifying the correct implementation of the safety requirements of such systems. A safety case is a form of assurance case. A safety case consists of a structured argument supported by a body of evidence. It aims at demonstrating that a given system is acceptably safe to operate in a given environment. It is represented either as plain text (e.g., structured prose, Trust case) or using graphical notations \cite{b22}. These graphical notations include GSN (Goal Structuring Notation) and CAE (Claim-Argument-Evidence) \cite{b18, b22}. Still, GSN is the most popular graphical notation to represent safety cases.

A Large Language Model (LLM) is a type of artificial intelligence (AI) model that is typically trained on extensive and diverse datasets. These models find utility across a wide range of applications, such as natural language processing (NLP), answering questions, and even generating code, among others. In the field of LLMs, OpenAI's GPT-4 \cite{b16} has attracted increasing attention in recent years and is recognized as one of the most influential LLMs. While GPT-4 has found applications in various software engineering tasks, its performance in the context of safety case modeling remains relatively unexplored. However, the utilization of generative AI models like GPT-4 for software modeling, particularly safety case modeling, may pose some challenges. Firstly, there exists a concern regarding the potential for hallucinated responses generated by GPT-4 \cite{b5, b23}. This implies that the safety arguments crafted by the model could contain inaccuracies or even false information. In situations involving high-risk critical systems (e.g., autonomous cars), such inaccuracies could lead to profoundly adverse consequences. Secondly, GPT-4 operates in a non-deterministic manner. In practical terms, this means that it has the capacity to provide varying responses to the same queries during different runs. This non-determinism adds an additional layer of complexity and unpredictability to the process of generating safety arguments using GPT-4.

Safety cases are usually very large documents \cite{b8}.
Still, they are usually generated manually, which can be a very tedious, error-prone, and time-consuming process, especially for complex and heterogeneous systems such as cyber-physical systems (CPS) \cite{b9,b10,b11}. It is therefore crucial to devise new techniques to automatically generate safety cases. 

In the realm of safety case research, despite the extensive body of work dedicated to automating the generation of safety cases, our investigation, to the best of our knowledge, stands as a pioneering effort in employing GPT-4 for this purpose. Consequently, the primary focus of our paper revolves around the automation of safety case creation through the utilization of GPT-4. The main objective of this endeavor is to alleviate the workload of safety case modelers, offering a novel approach to safety case development.

By adapting the methodology from Chen et al. \cite{b5} to safety cases, we conducted two sets of experiments. The first set aimed to assess GPT-4's proficiency in responding to fundamental queries related to the goal structuring notation (GSN), a notation for representing safety cases. In the second set, we delved into GPT-4's potential in generating safety cases. To achieve this, we employed two safety cases as ground-truths and carried out four different experiments, each involving variations in domain knowledge and GSN syntax understanding, to evaluate GPT-4's capability to generate safety arguments effectively.

The contributions of our work are as follows:
\begin{itemize}
    \item \textbf{Contribution 1}: We have extracted and elucidated the intricate structural and semantic rules that the GSN standard embodies
    \item \textbf{Contribution 2}: Furthermore, our work includes the development of both rule-based and generation-based questions as means to assess GPT-4's competence in understanding the GSN and generating GSN elements
    \item \textbf{Contribution 3}: We have  evaluated GPT-4's performance, focusing on its structural correctness, semantic accuracy, and overall reasonability. For this purpose, we have used GPT-4 to generate 32 safety cases similar to the two ground-truth safety cases at hand.
\end{itemize}
Our study provides a preliminary assessment of GPT and sheds light on the capabilities and limitations of GPT-4 in the realm of safety cases. Our work therefore contributes valuable insights to the research field of safety assurance coupled with AI.

The remainder of this paper is organized as follows. Section \ref{sec2} explains background concepts. Section \ref{sec3} presents related work. Section \ref{sec4} describes our methodology. Section \ref{sec5} reports the results. In Section \ref{sec6}, we discuss our results. Section \ref{sec7} presents the threats to validity and finally Section \ref{sec8} concludes and outlines future work.

\section{Background} \label{sec2}
\subsection{What is safety assurance?}

Ensuring the technological progress of cyber-physical systems (CPSs) can save lives, prevent injuries, reduce traffic and costs associated with car accidents, and the environmental impact of vehicles \cite{b15}. However, prior to their deployment, CPSs must be proved safe \cite{b8}. To achieve this, CPSs manufacturers usually develop compelling safety cases to demonstrate their CPSs sufficiently support the desired safety requirements, which allows regulatory bodies to certify the CPSs \cite{b14}. To prevent system failure, safety cases allow verifying the correctness of systems’ capabilities in compliance with specific industrial standards (e.g., ISO 26262, DO-178C) \cite{b8,b14}. A safety case is a safety assurance model, containing a hierarchy of claims, arguments and evidence that aims at demonstrating that a given system is sufficiently safe \cite{b1}.

Safety assurance consists of several activities including: safety cases design, representation, generation, automation, assessment and formal verification and validation, among others. Our focus lies in the domain of safety case generation, primarily due to the sheer complexity and size of safety cases, often exceeding 500 pages in length \cite{b8}. The manual creation of such extensive safety cases can be a daunting, time-consuming, and error-prone task \cite{b8}. Hence, it is imperative to propose novel techniques for (semi-)automatically generating safety cases.

\subsection{The Goal Structuring Notation and its structured prose}

A safety case can be presented in several ways including traditional prose (or simply plain prose), semi-structured text (also called structured prose) or graphical notations. Initially, safety cases were represented using traditional prose. However, presenting safety arguments in natural language may lead to unclear, unformal, and sometimes incomprehensible safety cases \cite{b17, b20}. An improvement over plain prose is called the structured prose. Contrary to plain prose, structured prose can be used to formally show the relationships between the elements of a safety case \cite{b17, b20}. Compared to plain and structured prose, graphical notations provide a coherent representation of safety arguments with their supporting evidence. In terms of graphical notations, various notations such as the goal structuring notation (GSN) and claim-argument-evidence (CAE) \cite{b18, b22} have been proposed to formally represent a safety case. GSN and CAE are aligned with the structured assurance case metamodel (SACM). The latter is a standard that the Object Management Group (OMG) introduced to promote interoperability and standardization \cite{b21}. 

GSN represents a safety case as a tree-like structure called \textit{goal structure} \cite{b35}. The core GSN elements that can be used to create a safety case represented as a goal structure are \cite{b18}:

\begin{itemize}
    \item \textbf{\textit{Goal:}} Presents a claim. The latter explains what the safety case is trying to demonstrate
    \item  \textbf{\textbf{Strategy:}} Embodies the inference that exists between a goal and its sub-goals
    \item \textbf{\textit{Solution:}} Represents the evidence
    \item \textbf{\textit{Context:}} Reference to contextual information
    \item \textbf{\textit{Assumption:}} This statement explains the assumption regarding how the supporting element reinforces the parent element
    \item \textbf{\textit{Justification:}} it provides a rationale for the inclusion or wording of a GSN element without altering the intended meaning of the element itself.
\end{itemize}    

The main GSN relationships allowing to connect GSN elements include the following \cite{b18}:
\begin{itemize}
    \item\textbf{\textit{InContextOf:}} Documents contextual relationships between GSN elements
    \item\textbf{\textit{SupportedBy:}} Documents supporting relationships between GSN elements
 \end{itemize}   

GSN elements can be decorated by relying on decorators. These include the following \cite{b18}:
 \begin{itemize}   
    \item\textbf{\textit{Undeveloped:}} Decorator indicating that the GSN element is yet to be developed

    \item\textbf{\textit{Uninstantiated:}} Decorator indicating that the GSN element is yet to be instantiated (i.e., the element needs to be replaced by a more concrete instance)
    \item\textbf{\textit{Off-Diagram:}} Decorator denoting that the GSN element continues in or continued from a separate diagram.
\end{itemize}

\begin{figure*}[t!]
  \centering
  \includegraphics[width=\linewidth]{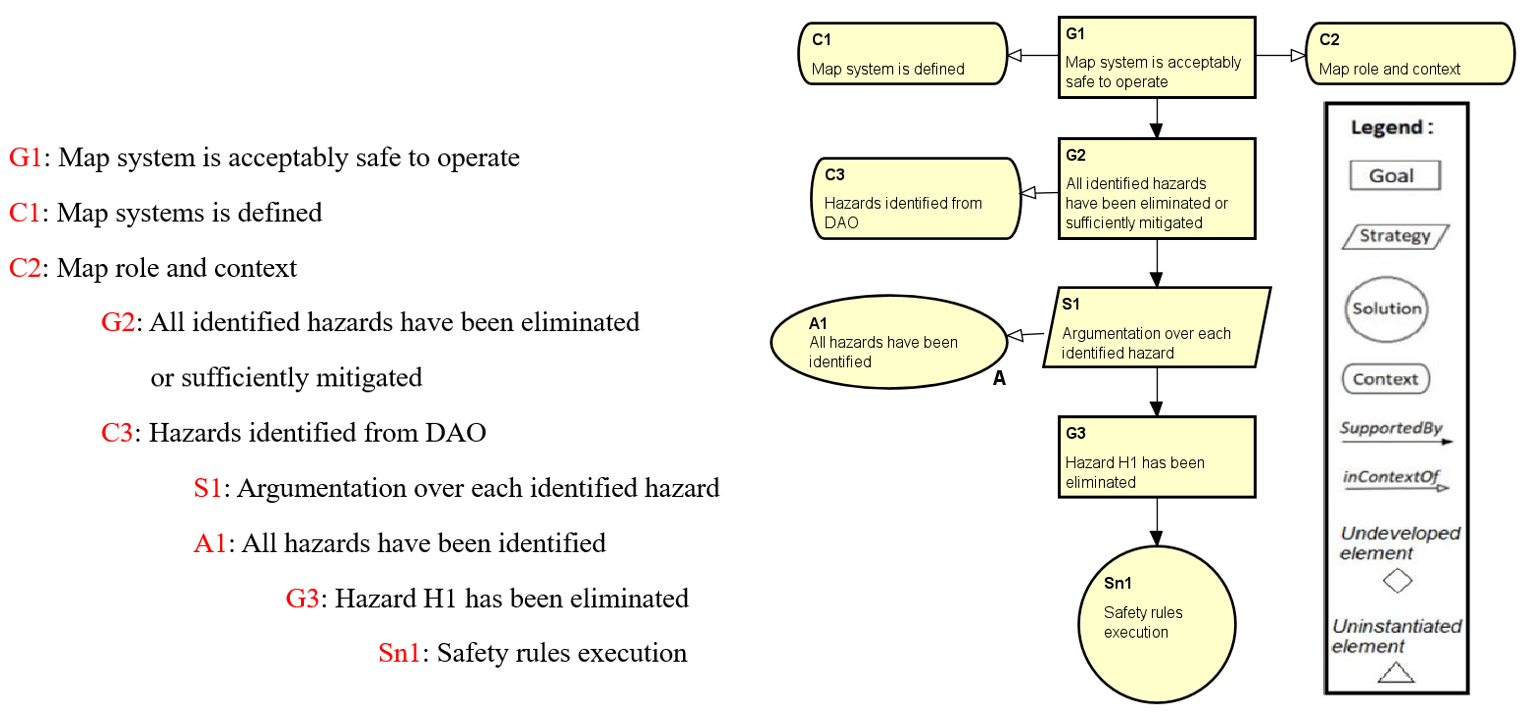}
  \caption{On the right, an example of safety case adapted from \cite{b19} and depicted in the GSN; on the left, the equivalent of the safety case in the structured prose.}
  \Description{This figure is an example of a safety case which is adopted from the reference \cite{b19}. It shows 2 kinds of safety case representation. structured prose and GSN}
  \label{fig1}
\end{figure*}

Figure \ref{fig1} is adapted from \cite{b19}. It provides an example of a safety case in GSN together with its structured prose equivalent. In that figure, the main objective of the safety case is to show that the map system is acceptably safe to operate, which is depicted by G1. G1 is defined in two contexts, C1, which corresponds to the definition of the map system and C2, which defines the role and context of the map. G1 is further supported by goal G2, which states that all the identified hazards are either eliminated or they have been sufficiently mitigated for safe operation. G3 is in the context of C3, which states that the hazards in G3 are the hazards identified from DAO (Dysfunctional Analysis Ontology). G2 is further supported by strategy S2 which places an argument over each of the identified hazard. It is also in the context of an Assumption A1 which states that all hazards have been identified. Finally, G3 shows that hazard H1 has been eliminated which is supported by the evidence Sn1, execution of the safety rules. Hence, through a series of goals, strategies, assumption, contexts and evidence, this safety case is able to argue that the map system is sufficiently safe to operate. 

In this paper, we chose to represent safety cases using the structured prose notation instead of the graphical notation. This better aligns with the GPT-4 capability of emitting text outputs. Still, structured prose safety cases comply with the GSN and can be turned into graphical GSN diagrams (i.e. goal structures) following the transformation guidelines outlined in the GSN standard version 3 \cite{b18}. 

\subsection{What is a large language model?}

A large language model (LLM) is a type of artificial intelligence (AI) model that have garnered attention in natural language processing tasks (NLP) recently. LLMs are usually transformer models (e.g., GPT-4) that are trained on massive data sets to generate text and answer questions with high accuracy. LLMs are typically characterized by their vast size and complexity, as they are trained on massive data sets that contain text from a variety of sources, including books, articles, websites, and more. These models use deep learning techniques, particularly neural networks, to process and generate text, and they are known for their ability to capture complex patterns and semantics in language \cite{b24}. Prominent examples of LLMs include OpenAI's GPT (Generative Pre-trained Transformer) series, such as GPT-3.5 and GPT-4 \cite{b16}, as well as models from other organizations like BERT (Bidirectional Encoder Representations from Transformers) \cite{b25} developed by Google. In this paper, our focus is on GPT-4.0 since it is more powerful than most of the LLMs present in the market and it is easily accessible.

\subsection{What are the prompting techniques that have been proposed for LLMs?}

In literature, various prompting techniques have been introduced to enhance the quality of responses generated by ChatGPT and to maximize the extraction of information with a strong emphasis on precision. For instance, Wang et al. \cite{b25} have employed prompting techniques in their work on VOYAGER, an embodied lifelong learning agent in Minecraft. VOYAGER interacts with GPT-4 through blackbox queries, eliminating the need for model parameter fine-tuning. They use a prompting method where they ask ChatGPT for guidance on the next task to be performed in Minecraft. To provide context about the player's current situation in the game, they engage in a question-and-answer exchange with GPT. They also specify the desired format for GPT-4's responses as part of their instructions. The work by Chen et al. \cite{b5} utilize prompting in their experiments to gauge the ability of GPT-4 in generating goal models in goal-oriented requirement language. They employ various prompting techniques such as single sentence, single sentence with domain paragraph, syntax description + single sentence and finally syntax description + single sentence + domain paragraph. In this paper, we embrace and build upon the prompting methods suggested in these previous studies.

\section{Related work} \label{sec3}
\subsection{Automatic generation of assurance cases}
In the last decade, many researchers have actively contributed to the automatic generation of safety cases. For example, Denney and Pai \cite{b1} propose a methodology that automatically generates safety cases derived from a formal verification method. They have applied their methodology to the Swift Unmanned Aircraft System (UAS) which is being developed at NASA. Nonetheless, the research lacks a method for evaluating or assessing the confidence level of the automatically generated assurance cases.

Similarly, Wang et al. \cite{b2} present an automated assurance case generation framework that helps in creating, validating and assessing safety cases. While their work represents a significant contribution to automating safety cases, it is worth noting that in their evaluation, they focus exclusively on specific sections of the assurance cases that are either fully complete or nearly complete. This approach may not provide a completely accurate assessment. 

Ramakrishna et al. \cite{b3} argue that manual methods of creating safety cases are very brittle and do not have a systematic approach. As an improvement to current practises, the authors introduce a tool named structured ACG (assurance case generation) tool which uses design artifacts, evidence and expertise of developers to construct a safety case automatically. The study's primary limitations are the need for human input in converting certification criteria into GSN goals and in evaluating safety cases with just a single metric, and the lack of automation in linking evidence to system components and defining assumptions.

In our paper, we address these identified limitations by proposing a method that utilizes GPT-4 for the automatic generation of  safety cases in structured prose, aligned with the Goal Structuring Notation (GSN). Additionally, we introduce a comprehensive evaluation of these safety cases using three distinct measures.

\subsection{On the use of large language models to automatically generate software models}
In recent times, there has been a notable increase in the application of LLMs to tackle various software engineering tasks. These tasks span a wide range, including the automated recommendation of design concepts for metamodel development, the generation of goal models, and even the creation of UML (Unified Modeling Language) models. This surge in interest reflects the growing recognition of the potential of LLMs to enhance and automate different aspects of the software engineering process. 

For instance, in \cite{b4}, the authors discuss the challenge of designing metamodels, which define complex relationships between concepts in Model-Driven Engineering (MDE). They propose an approach that uses Deep Learning --and more specifically pre-trained language models-- to recommend relevant domain concepts for metamodel designers. By training a model on a large dataset of metamodels to learn from both structural and lexical properties, the authors were able to show that the model can accurately provide recommendations for concept renaming scenarios.

More recently, Chen et al. \cite{b5} investigate the potential of advanced language models like GPT-4 in the field of requirements engineering, particularly in the development of goal-oriented models. The study focuses on GPT-4's familiarity with a specific modeling language called the Goal-oriented Requirement Language (GRL). The results indicate that GPT-4 possesses substantial knowledge about goal modeling. While some generated elements were generic or incorrect, the generated concepts added value, especially for stakeholders unfamiliar with the domain. In this paper, we adapt and extend on this research.

Besides, Chaaben et al. \cite{b6} and Camara et al. \cite{b7} have also used ChatGPT to generate UML models. The former introduces a novel approach to enhance domain modeling using LLMs without extensive training. The approach leverages few-shot prompt learning and doesn't require vast datasets for model training. While the latter discusses the potential impact of LLMs like ChatGPT, focusing on their capabilities and constraints when it comes to software modeling tasks, especially in the context of UML modeling.


Viger et al. \cite{b12}  proposed to use generative AI to elicit defeaters i.e. potential doubts in the arguments. They proposed to use GPT-4 to aid in the identification of defeaters in assurance cases to make them more reliable. That work is still preliminary, so the authors did not empirically validate it. Thus, to the best of our knowledge, the use of ChatGPT to generate safety cases has never been explored before. Hence, in this paper, we address that gap by exploring the use of LLMs, specifically, GPT-4, to generate safety cases.

\section{Methodology} \label{sec4}

\subsection{Research questions}
Through our preliminary experiments, we aim to answer the following research questions (RQs):

\textbf{RQ1: Is ChatGPT sufficiently proficient in the Goal Structuring Notation?} Similar to Chen et al. \cite{b5}, we will employ a set of 19 questions to answer our research inquiry. These questions are designed to evaluate GPT-4's comprehension of the structural and semantic rules that define the Goal Structuring Notation (GSN) and its ability to generate concise goal structures. 

\smallskip

\textbf{RQ2: Is ChatGPT able to generate safety cases that are structurally, semantically, and reasonably correct?}

To address our second research question, we will utilize two safety cases from the literature as our ground-truths. These safety cases pertain to different application domains: one in the field of energy and the other in the automotive sector.

\subsection{GPT-4 setting}
To address our research questions (RQs), we have utilized the online web platform for ChatGPT, specifically the premium version of GPT-4, available at \url{https://chat.openai.com/}. This platform is instrumental in running our prompts and retrieving GPT-4 answers.

\subsection{How to prompt GPT-4 to answer RQ1?}

\subsubsection{Extraction of structural and semantic rules from GSN standard v.3}
To document the structural and semantic rules governing the Goal Structuring Notation (GSN), we referenced the GSN standard v.3 document \cite{b18}. That document has been created and is maintained by the Goal Structuring Notation Working Group, with the objective of disseminating information and resources related to GSN \cite{b18}. We systematically analyzed that document to extract and derive the structural and semantic rules that the GSN embodies. This analytical process involved a detailed examination of the standard's content. The rule extraction plays a pivotal role in our research as it provides the essential groundwork for formulating fundamental GSN-related questions. Table \ref{tab:rules} outlines these extracted rules. Structural rules pertain to the arrangement and configuration of GSN elements, specifying how each element should be structured and the connections they should have with other elements. 

The rules that Table \ref{tab:rules} reports define the hierarchy and relationships within GSN, ensuring that the notation is used consistently and effectively. On the other hand, semantic rules focus on the textual content and meaning (semantic) contained within GSN elements. These rules encompass aspects such as grammar, terminology, and the interpretation of the textual information specified in each GSN element, ensuring that the notation conveys information accurately and unambiguously.

\begin{table*}
  \caption{GSN Structural and semantic rules}
  \label{tab:rules}
  \begin{tabular}{p{2cm} p{9cm} p{4cm}}
    \toprule
    \textbf{GSN element} & \textbf{Structural rules} & \textbf{Semantic rules} \\
    \midrule
    Goal & Allowed connections: Goal to goal, Goal to strategy, Goal to solution, Goal to context, Goal to assumption, Goal to justification & Noun phrase + verb phrase\\
    Context & Allowed connections: Goal to context, Strategy to context & Noun phrase\\
    Strategy & Allowed connections: Strategy to goal, Strategy to context, Strategy to assumption, Strategy to justification & Brief description of the argument approach \\
    Solution & Allowed connections: Goal to solution & Noun phrase\\
    Justification & Allowed connections: Goal to justification, strategy to justification & Noun phrase + verb phrase\\
    Assumption & Allowed connections: Goal to assumption, strategy to assumption & Noun phrase + verb phrase\\
    \bottomrule
  \end{tabular}
\end{table*}

\subsubsection{Description of the questions asked to GPT-4}

Table \ref{tab:qns} presents the sample questions which we framed for evaluating GPT-4's proficiency on GSN. As detailed in the preceding section, we devised a set of 19 questions, similar to Chen et al. \cite{b5}, to gauge GPT-4's competence in responding to inquiries related to Goal Structuring Notation (GSN), drawing inspiration from the GSN standard v.3 document \cite{b18}. Our assessment encompasses the following two categories of questions. 
\begin{enumerate}
    \item \textbf{\textit{Rule-based questions:}} we created an initial set of 13 questions, each addressing either the structural or semantic rules conveyed by the GSN standard
    \item \textbf{\textit{Generation-based questions:}}  In addition to the initial set of 13 questions, we introduced 6 generation-based questions to evaluate GPT-4's capability to generate fundamental GSN elements. These questions assess GPT-4's aptitude in generating core components of GSN. For the sake of brevity, Table \ref{tab:qns} only reports a representative question for each category. The complete list of 19 questions is available online \footnote{\url{https://github.com/AnonymousAuthours/Safety_Cases_GPT4}}.
\end{enumerate}

\begin{table*}
  \caption{Sample inquiry questions for GPT-4 regarding goal structuring notation (GSN)}
  \label{tab:qns}
  \begin{tabular}{p{3cm} p{5cm} p{9cm}}
    \toprule
    Category of Question & Sub-category & Sample Question \\
    \midrule
    Rule-based & Structural-correctness based Question & How many elements are present in a goal-structure and what are they? Can a parent element have multiple children? \\ [0.25cm]
               & Semantic-correctness based questions & Explain what a top-level claim is. Can it be supported by multiple sub-claims? \\ [0.25cm]
    Generation-based & -  & Give me a sample goal element connected to 2 sub-goals \\
    \bottomrule
  \end{tabular}
\end{table*}

\subsubsection{Description of the prompt structures used for RQ1}

Before asking the 19 questions to GPT-4, we took a preparatory step by employing specific prompting techniques outlined in \cite{b26} and \cite{b30}. The primary objective of this preparatory prompt (presented in red in the text box below) was to orient GPT-4 appropriately, ensuring that it understood its role as an assistant in helping us address inquiries regarding GSN. Furthermore, we considered the tendency of GPT-4 to provide lengthy responses. To mitigate this, we included a directive statement requesting GPT-4 to provide concise answers. This was crucial to maintain the clarity and brevity of the responses. Additionally, we took into account that, without explicitly mentioning "GSN" as an abbreviation for goal structuring notation, GPT-4 might interpret it differently, such as "~\textit{Game Show Network}". To avoid ambiguity and ensure precision, we followed the approach established by Chen et al. \cite{b5}, where setting a context becomes essential. This context-setting step ensured that GPT-4 would respond accurately and in alignment with our intended subject matter, which is GSN. The preparatory prompt is given below:

\smallskip
\noindent\fbox{%
    \parbox{\linewidth}{%

\textcolor{red}{You are an assistant that helps me answer questions about Goal structuring notation (GSN). GSN always refers to Goal Structuring Notation from this point. Your answers should be concise and to the point. It should not be more than 2-3 lines}

}
}

\subsection{How to prompt GPT-4 to answer RQ2?}

\subsubsection{Description of safety cases used as ground-truths}

To evaluate both the structural accuracy and semantic correctness of the safety cases generated by ChatGPT, we will use established reference safety cases as benchmarks. These safety cases respectively focus on the energy and automotive application domains. These two reference safety cases consist of the X-ray safety case outlined in \cite{b27} and the safety case related to a Machine-Learning (ML) enabled component for a tire noise recognition (TNR) in a vehicle, is documented in \cite{b28}. The X-ray system described in \cite{b27} bears similarities to airport X-ray backscattering machines. It qualifies as a safety-critical system due to the potential harm to individuals in proximity if the radiation emitted by the X-ray machine exceeds acceptable limits. 

Similarly, in \cite{b28}, the authors present an assurance case approach for a TNR component that employs machine learning (ML) to classify road conditions based on audio signals from wheel-mounted microphones. This ML algorithm is vital for safety because it informs chassis control and powertrain systems to adapt control parameters, ensuring consistent traction, and the system's real-time processing is crucial for accurate road surface assessment, especially in critical situations. 

\begin{figure*}[t!]
  \centering
  \includegraphics[width=\linewidth]{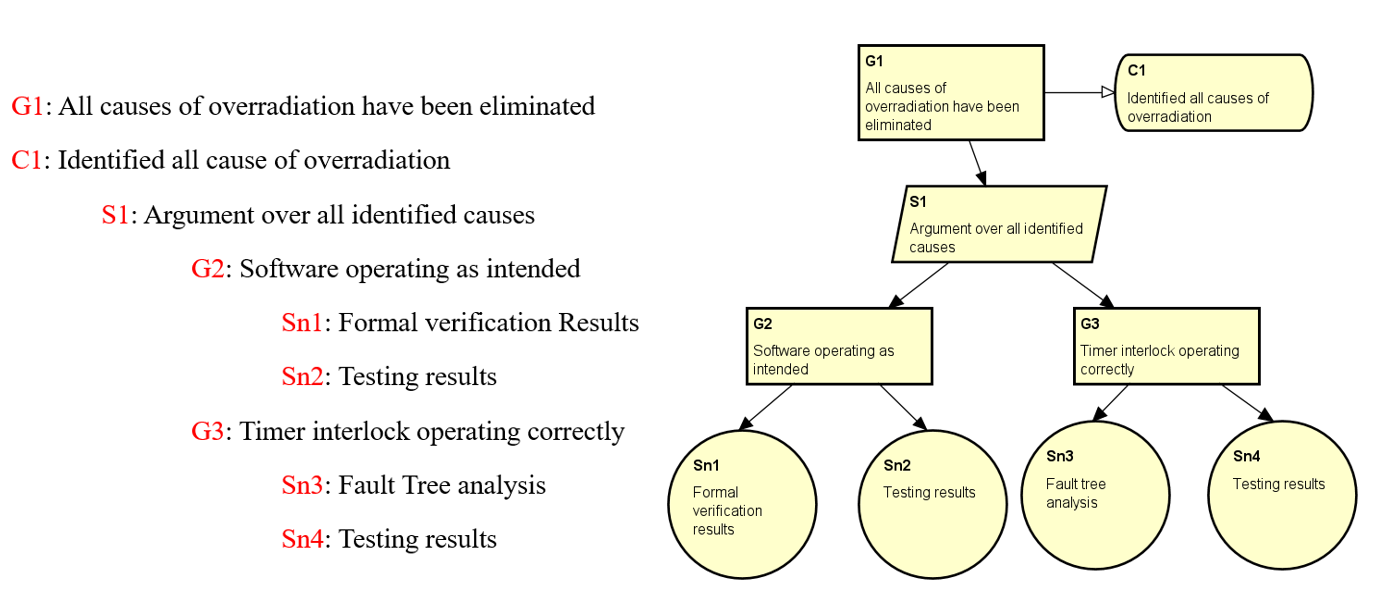}
  \caption{On the right, the X-ray safety case adapted from \cite{b27} and depicted in the GSN; on the left, the equivalent of the safety case in the structured prose.}
  \Description{This figure is an example of a safety case which is adopted from the reference \cite{b27}. It shows 2 kinds of safety case representation. structured prose and GSN}
  \label{fig2}
\end{figure*}

\begin{figure*}[t!]
  \centering
  \includegraphics[width=\linewidth]{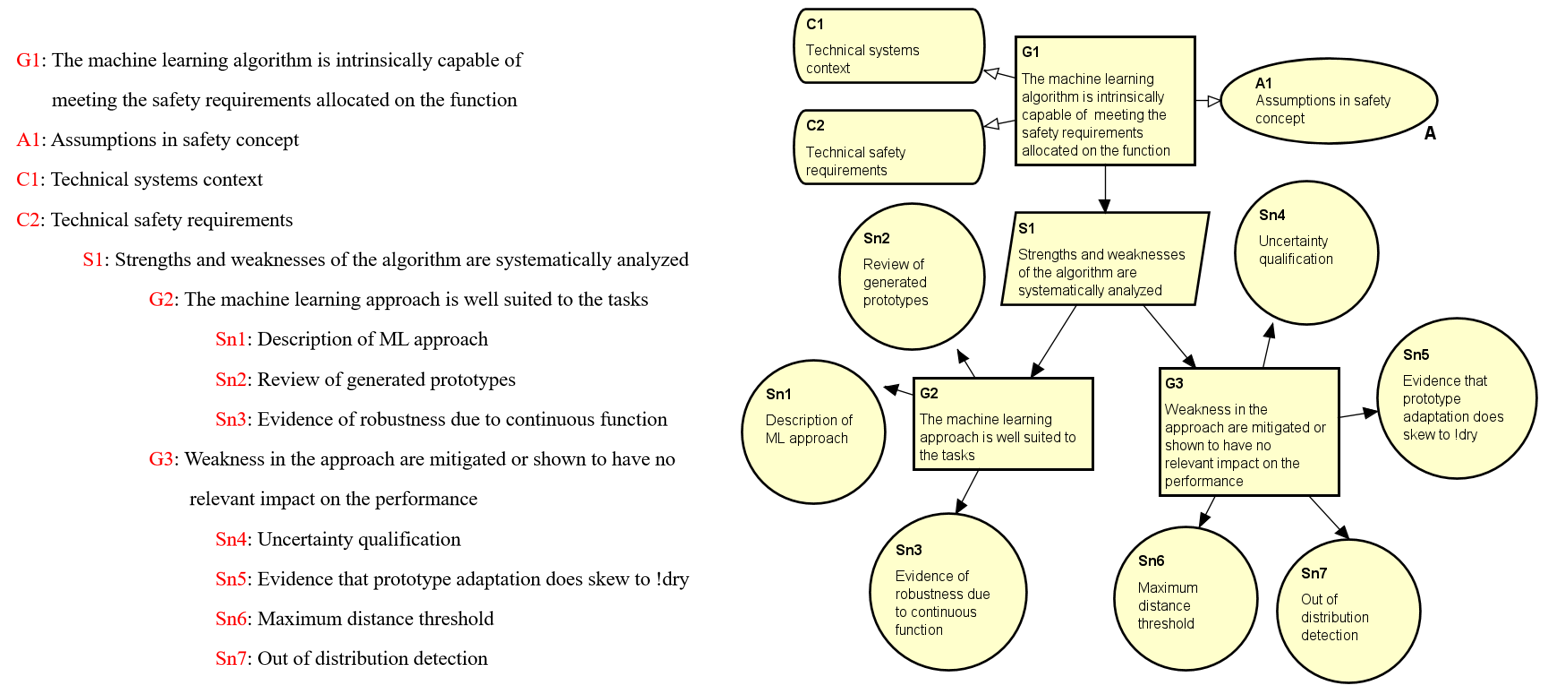}
  \caption{On the right, the ML safety case adapted from \cite{b28} and depicted in the GSN; on the left, the equivalent of the safety case in the structured prose.}
  \Description{This figure is an example of a safety case which is adopted from the reference \cite{b28}. It shows 2 kinds of safety case representation. structured prose and GSN}
  \label{fig3}
\end{figure*}

In Figure \ref{fig2}, we provide a visual representation of the X-ray safety case in both Goal Structuring Notation (GSN) and its equivalent in the structured prose. Figure \ref{fig3}, on the other hand, presents a similar representation of the ML algorithm safety case. The primary objective of the X-ray safety case is to demonstrate the elimination of all factors leading to overradiation (G1). This is substantiated by a strategy (S1) and two subsidiary goals (G2 and G3), along with supporting pieces of evidence (i.e. Sn1, Sn2, Sn3, Sn4), all working together to validate the overarching claim. Similarly, the primary aim of the ML algorithm safety case is to confirm that the algorithm complies with the safety requirements allocated to its function (G1). This top goal is reinforced by a strategy (S1), which is further subdivided into two sub-goals (G2 and G3) and substantiated by seven pieces of evidence (Sn1-Sn7).

\subsubsection{Description of experiments ran to answer RQ2}
To evaluate ChatGPT's ability to generate concrete safety cases, we will adapt the experiments that  Chen et al. performed on goal models \cite{b5}. For each of the two established safety cases (considered as ground-truths), we will carry out four experiments. In these experiments, we will task ChatGPT (specifically, GPT-4) with automatically generating the same safety cases. The four experiments are explained as follows:
\begin{itemize}
    \item \textbf{\textit{Experiment 1:}} We provide instructions to generate the safety case but provide no domain knowledge, and we do not specify GSN syntax (i.e., no GSN structural rule is specified, and no example of GSN syntax is provided)
    \item \textbf{\textit{Experiment 2:}} We provide instructions to generate the safety case, and we provide domain knowledge, and we do not specify GSN syntax (i.e., no GSN structural rule is specified, and no example of GSN syntax is provided). 
    \item \textbf{\textit{Experiment 3:}} We provide instructions to generate the safety case, and we do not provide domain knowledge, but we specify GSN syntax (GSN structural rules are specified, and an example of GSN syntax is provided). 
    \item \textbf{\textit{Experiment 4:}} We provide instructions to generate the safety case, and we provide domain knowledge, and we specify GSN syntax (GSN structural rules are specified, and an example of GSN syntax is provided). 
\end{itemize}

We ran each of these four experiments \textbf{\textit{k}} times, choosing k value as 4, following a similar approach as Chen et al. \cite{b5}. This choice of conducting multiple rounds is motivated by the non-deterministic nature of GPT-4. It helps mitigate potential variability in responses and ensures a more robust and comprehensive evaluation of GPT-4.

\subsubsection{Description of the prompt templates used for RQ2 }

In our approach to framing the prompting templates for RQ2, similar to RQ1, we initiated the process with a preparatory prompt to inform GPT-4 of its role as a safety case developer assistant. Subsequently, for all four experiments, which encompass scenarios with and without context, and with and without GSN syntax, we adopted the prompting techniques outlined in \cite{b26} as our foundation. In the referenced paper, the authors offer prompts structured in a question-and-answer format to provide additional context to GPT-4. Building upon this methodology, we tailored our prompts to suit our platform's requirements. These prompts included sets of questions and corresponding answers designed to elucidate the concept of a safety case and the specific format in which we expected GPT-4 to generate the safety case. For experiments involving additional domain context, we extended this approach by incorporating further information in the form of question-and-answer pairs. An example of the prompt template utilized for Experiment-1 (lacking context and GSN syntax) in the context of the ML algorithm safety case (the one in Figure \ref{fig3}) is provided in the text box below:

\smallskip
\noindent\fbox{%
    \parbox{\linewidth}{%

\textbf{Prompt:}

\textcolor{red}{You are a professional safety case developer assistant. I want you to create a safety case for the given system in Goal Structuring Notation (GSN) Format.}

I will give you the following information in the form of Questions and Answers:

\textbf{Question 1: What is a safety case?}
Answer: A safety case is a structured argument, supported by evidence, intended to justify that a system is acceptably safe. 
\newline
\textbf{Question 2: What is the format of the safety case}
Answer: I want you to generate a safety case in GSN Format
\newline
\textbf{Question 3: What is the system for which you need to generate a safety case}
Answer: The system is a Machine Learning (ML) algorithm that is used to implement the classification function of a Tire Noise Recognition (TNR) component of a vehicle.
\newline
\textbf{Question 4: What is the main objective of the safety case}
Answer: The objective of the safety case is to develop a structured and convincing argument that the classifier fulfilled its technical requirements, with respect to functional inefficiencies that could lead to False Positives (FP) identifications of dry road surface conditions.
\newline
\textcolor{red}{Create a top-level safety case for the ML algorithm in GSN format.}

}
}
\smallskip

\subsection{Measures used to assess the results}\label{sec45}

To answer RQ1, we conducted two distinct types of assessment. Firstly, to assess GPT-4's responses to each of these 19 questions, we will enlist the expertise of two graduate students, all well-versed in GSN and its associated structured prose. Each researcher will independently score the answers on a linear scale ranging from 1 (completely accurate) to 5 (inaccurate). Subsequently, we will calculate 
the Kendall's Tau correlation \cite{b31}.similar to Chen et al. \cite{b5}, to determine the inter-rater agreement. We utilized an online calculator for this purpose\footnote{\url{https://www.gigacalculator.com/calculators/correlation-coefficient-calculator.php}}.
The value of the Kendall's Tau progresses from 0 to 1. When the value of the Kendall's Tau correlation is close to 0, this means that the level of agreement is almost close to none. However, when the value is close to 1, then the level of agreement is strong. Equation (\ref{eq1}) outlines the computation of Kendall's Tau.

    \begin{equation}\label{eq1}
    \tau = \frac{C - D}{C + D}
\end{equation}
\smallskip

In that equation, \textit{C} is the number of concordant pairs, while \textit{D} is the number of discordant pairs.

Secondly, as part of the second assessment, we sought to determine whether GPT-4 could attain a passing grade concerning its knowledge of the goal structuring notation (GSN). To achieve this, we computed the average scores across the four rounds \textbf{(k=4)} to establish the grade achieved by GPT-4.

Like Chen et al. \cite{b5}, in RQ2, the assessment is done by three assessors (two graduate students and a faculty member), and meetings were held to resolve any potential conflicts in the assessment process. To answer RQ2,  we use GPT-4 to generate safety cases for each of the 4 experiments. Each experiment consists of \textbf{k=4} rounds (runs), and GPT-4 generates one safety case per round. GPT-4 therefore generates a total of sixteen safety cases for each of the two ground-truth safety cases used in the four experiments. We relied on the following three measures for the assessment of the safety cases that GPT-4 generated when we performed the 4 experiments: 
\begin{enumerate}
    \item \textbf{\textit{Ground-Truth similarity}}: This measure assesses the resemblance between the safety case generated by GPT-4 and the ground-truth safety case. To assess the ground-truth similarity, each safety case generated by GPT-4 in the structured prose is visually represented in the GSN. The resulting goal structure is rated by three raters against the ground-truth safety case represented in the GSN. Ratings are based on the following linear scale: \textbf{\textit{1=Totally correct, 2=Mostly correct, 3=Moderately correct, 4=Slightly correct, 5=Incorrect}}.
    \item \textbf{\textit{Reasonability}}: Reasonable means the GSN “element could reasonably be in the ground-truth but is not” \cite{b5}. To assess the reasonability, each safety case generated by GPT-4 in the structured prose is first visually represented in the GSN. The resulting goal structure is rated by three raters based on the following linear scale: \textbf{\textit{1=Totally reasonable; 2=Mostly reasonable; 3=Moderately reasonable; 4=Slightly reasonable; 5=Unreasonable}}.
    \item \textbf{\textit{Semantic similarity:}} Here, we rely on the cosine similarity to automatically compute the semantic similarity (resemblance) between the ground-truth of a safety case (in the structured prose) and each safety case generated by ChatGPT and represented in the structured prose. The value of cosine similarity is between -1 and 1 with -1 being no similarity and 1 being similar. To compute cosine similarity automatically, we utilized python scripts. 
    \end{enumerate}

Noteworthy, to compute the cosine similarity, our Python program makes use of the BERT tokenizer and BERT model available in the Transformers library. It takes two input text files: one containing the ground-truth safety case in structured prose, and the other containing the GPT-4 generated safety case in structured prose. Initially, these text files are processed and tokenized by the pre-trained BERT model, enabling the computation of cosine similarity between them. The cosine similarity function, sourced from the sklearn library, is employed to quantify the similarity score between the two documents. The rationale behind using the pre-trained BERT model is its extensive training on massive datasets, leading to its remarkable performance and state-of-the-art results across a spectrum of natural language processing tasks \cite{b34}.

\subsection{Supplemental material availability}
The supplemental material containing the python program used to compute cosine similarity, the complete prompts that we used to perfom experiments aiming at assessing both RQ1 and RQ2, is available online \footnote{\url{https://github.com/AnonymousAuthours/Safety_Cases_GPT4}}.

\section{Results} \label{sec5}

\subsection{RQ1: is ChatGPT sufficiently proficient in the Goal Structuring Notation?}

As detailed in Section \ref{sec45}, our evaluation process encompassed two distinct types of assessments. In the first assessment, where we had two graduate students (assessors/raters) grade the responses to the 19 questions posed to GPT, we aimed to quantify the inter-rater agreement using the Kendall's Tau correlation. The results of this analysis are presented in Table \ref{tab:avgkendall}. These results indicate the average correlation between the scores assigned by the two raters is \textbf{73\%}. This high level of correlation signifies a very strong agreement between the two raters assessments.

\begin{table}
  \caption{Value of kendalls' Tau correlation for each of the four 4 runs of RQ1 along with its average}
  \label{tab:avgkendall}
  \begin{tabular}{ccccc}
    \toprule
   R1 & R2 & R3 & R4 & Avg\\
    \midrule
    0.66 & 0.86 & 0.63 & 0.77 & 0.73\\
    \bottomrule
  \end{tabular}
\end{table}

\begin{table}
  \caption{Average scores of all questions for RQ1}
  \label{tab:avgscore}
  \begin{tabular}{p{1cm}p{1cm}p{1cm}p{1cm}p{1cm}p{1cm}}
    \toprule
    Assessor & R1 & R2 & R3 & R4 & Avg \\
    \midrule
    1 & 1.37 & 1.63 & 1.53 & 1.37 & 1.48  \\
    2 & 1.16 & 1.58 & 1.42 & 1.47 & 1.40 \\
    \bottomrule
  \end{tabular}
\end{table}

Secondly, we aimed to evaluate GPT-4's competence in responding to fundamental queries related to Goal Structuring Notation (GSN). The results of this assessment, including the average scores from the four rounds of prompting GPT-4 to answer GSN-related questions, are presented in Table \ref{tab:avgscore}. Analysis of these results reveals that the average score across all four rounds for both the assessors is 1.44 (Avg of 1.48 and 1.40), indicative of GPT-4's robust understanding of the fundamental aspects of GSN. According to the grading scheme utilized by several universities, this performance corresponds to a \textbf{grade of A.} It is noteworthy that GPT-4 excelled in answering questions related to semantic correctness. Besides, GPT-4 is able to correctly specify how most of the GSN elements should be depicted and connected by relationships. GPT-4 is therefore able to generate safety cases in a hierarchical format, mirroring the structure of a goal hierarchy and resembling structured prose. In most of the rounds, it is also able to correctly generate samples of GSN elements. 

In the preceding inquiry, it is important to note that GPT-4 provided incorrect responses in relation to assumptions and justifications, which should have been represented using ellipses and not "angled brackets" or "square brackets" as GPT-4 suggested. The textbox below illustrates that incorrect answer (highlighted in red) that GPT-4 generated during our experiments.

\smallskip
\noindent\fbox{%
    \parbox{\linewidth}{%

\textbf{Qn: Can you explain how each of the six elements of GSN are structurally represented. i.e., what shapes?}

    Goals: Represented by rectangles.
    Strategies: Illustrated using parallelograms.
    Solutions: Shown as circles or ovals.
    Contexts: Depicted with rounded rectangles.
    \textcolor{red}{Assumptions: Represented by text in angled brackets.
    Justifications: Illustrated as text in square brackets.}

}
}
\smallskip

This indicates a limitation of GPT-4's understanding of the GSN syntax (notation) related to these specific two elements. Still, adopting the structured prose complying with the GSN instead of a visual representation prevents that limitation from having an impact on the GSN elements generated by GPT-4.
\smallskip

\noindent\fbox{%
    \parbox{\linewidth}{%

In summary, for \textbf{RQ1}, our conclusion is that GPT-4 has demonstrated a solid grasp of fundamental GSN knowledge, and it successfully passed this assessment with a grade of A.

}
}
\smallskip

\subsection{RQ2: is ChatGPT sufficiently able to generate a structurally, semantically and reasonably correct safety cases?}

As described in Section \ref{sec45}, our assessment of GPT-4's capacity to generate safety cases in response to RQ2 involved the participation of three assessors: two graduate students and one faculty member. The outcomes of their grading processes for GPT-4's responses were recorded for analysis. The results, which encompass average ground-truth similarity scores, average reasonability scores, and average cosine similarity scores, across four experiments and four rounds, are presented in Table \ref{tab:avggtrq2}, Table \ref{tab:avgrearq2}, and Table \ref{tab:cosine}, respectively. 

As shown in Table \ref{tab:avggtrq2}, in the case of the ML safety case, the average ground-truth similarity score across all four experiments amounts to 3, signifying that GPT-4 managed to produce safety cases that can be considered moderately accurate. Likewise, for the X-ray safety case, the average ground-truth similarity score comes out to be 3.38, which when rounded to the nearest integer also yields 3. This result suggests that, in the context of the X-ray safety case, GPT-4 was able to generate safety cases that are moderately accurate. As shown in Table \ref{tab:avggtrq2}, the average ground-truth similarity scores of 3 for both safety cases suggest that while GPT-4's generated safety cases that are moderately accurate, there is still room for improvement. 

The average reasonability scores (see Table \ref{tab:avgrearq2}), encompassing all four experiments, yield an average of 2.69. When rounded to the nearest whole number, this results in a score of 3. In the context of the ML safety case, this indicates that GPT-4's generated safety case elements are moderately reasonable. Similarly, for the X-ray safety case, the average reasonability score is 2.27, rounding off to a score of 2. As an example, in experiment-3 of the X-ray safety case, during run 1, GPT-4 generated an additional goal emphasizing \textbf{"Compliance with all relevant safety standards and regulations."} This goal is substantiated by a strategy of \textbf{"Argument based on compliance evidence,"} accompanied by two sub-goals: \textbf{"Machine design adheres to safety standards"} and \textbf{"Operational procedures align with safety regulations."} These sub-goals are further supported by evidence such as \textbf{"Compliance certification documents"} and \textbf{"Training records and operation manuals."} Despite not being part of the original safety case, these elements constitute valid GSN elements represented in the structured prose.

This suggests that, in the case of the X-ray safety case, GPT-4's generated elements are mostly reasonable. These findings suggest that GPT-4 demonstrates the capability to generate coherent safety case elements. While there may be instances where certain elements differ from the actual ground-truth safety case, they are generally considered to be moderately or mostly reasonable, depending on the specific safety case context. This underscores GPT-4's ability to generate safety case components that exhibit a reasonable level of coherence and logic. This implies that GPT-4 is able to generate GSN elements that the safety case modeler (argument developer) may not have thought of, but that can still be relevant for the modeled safety case. Hence, GPT-4 has the ability to generate GSN elements that can enrich the safety cases being modeled by a safety case modeler.

\begin{table}
  \caption{Average ground-truth similarity scores of 4 runs of each of the 4 experiments of RQ2 along with the total average}
  \label{tab:avggtrq2}
  \begin{tabular}{p{2cm}p{1cm}p{1cm}p{1cm}p{1cm}p{1cm}}
    \toprule
    Safety case & Exp-1 & Exp-2 & Exp-3 & Exp-4 & Avg\\
    \midrule
    ML & 3.5 & 2.8 & 3.25 & 2.5 & 3\\
    X-ray & 3.5 & 3 & 3.5 & 3.5 & 3.38\\
    \bottomrule
  \end{tabular}
\end{table}

\begin{table}
  \caption{Average reasonability scores of 4 runs of each of the 4 experiments of RQ2 along with the total average}
  \label{tab:avgrearq2}
  \begin{tabular}{p{2cm}p{1cm}p{1cm}p{1cm}p{1cm}p{1cm}p{1cm}p{1cm}}
    \toprule
    Safety Case & Exp-1 & Exp-2 & Exp-3 & Exp-4 & Avg\\
    \midrule
    ML & 2.75 & 2.75 & 2.5 & 2.75 & 2.69\\
    X-ray & 2.5 & 2 & 2.25 & 2.5 & 2.31\\
    \bottomrule
  \end{tabular}
\end{table}

\begin{table}
  \caption{Average cosine similarity scores of 4 runs of each of the 4 experiments of RQ2 along with the total average}
  \label{tab:cosine}
  \begin{tabular}{p{2cm}p{1cm}p{1cm}p{1cm}p{1cm}p{1cm}}
    \toprule
    Safety Case & Exp-1 & Exp-2 & Exp-3 & Exp-4 & Avg \\
    \midrule
    ML & 0.9	&0.88	&0.897	&0.895 & 0.89 \\
    X-ray & 0.867 & 0.902	&0.817 & 0.895 & 0.87 \\
    \bottomrule
  \end{tabular}
\end{table}

Finally, when examining the average cosine similarity scores that Table \ref{tab:cosine} reports, we find that for the ML safety case across all four experiments, the score is 0.89, and for the X-ray safety case, it is 0.87. These cosine similarity values are notably close to 1, which indicates a high level of similarity between the ground-truths and GPT-4 generated safety cases. From the perspective of cosine similarity, we can infer that GPT-4 successfully generated safety cases that are semantically similar to the original ground-truth safety cases. 

An interesting observation arises when we analyze these three measures together. The ground-truth similarity and reasonability measures both fall into the moderate range, However, the cosine similarity measure scores a high value. This contrast may stem from the fact that the computation of cosine similarity primarily considers the semantic resemblance between two safety cases, without directly assessing their structural correctness. Therefore, while GPT-4's generated safety cases exhibit semantic similarity to the reference cases, there may still be areas where their correctness or reasonability can be improved. These observations highlight the importance of considering multiple evaluation measures to comprehensively assess the quality of the generated safety cases.

\smallskip
\noindent\fbox{%
    \parbox{\linewidth}{%

In summary, our findings for \textbf{RQ2} indicate that GPT-4 possesses the capability to generate safety cases that are moderately reasonable and moderately correct. Additionally, GPT-4 generated safety cases exhibit a high degree of similarity with the ground-truth safety cases.

}
}
\smallskip

\section{Discussion}\label{sec6}

\subsection{Can ChatGPT alleviate the argumentation task of the safety case modeler?}

Our experiments show GPT-4 is able to generate safety arguments that are moderately accurate and reasonable when provided with relevant information. These moderate generation abilities led us to the conclusion that human intervention is still essential in the safety case generation process. There are two possible avenues for this intervention. The first one is incremental refinement, as established in Chen et al. \cite{b5}. Human involvement can take the form of incremental guidance, where a human interacts with GPT-4 to refine and shape the generated safety arguments. This iterative process ensures that the arguments align more closely with the specific system in question and adhere to the desired safety standards. The second one is manual expert analysis, where, a human safety expert can conduct a thorough manual analysis of the safety arguments generated by GPT-4. This expert can verify that all relevant information is captured and may make necessary changes to tailor the safety arguments to the particular system, thereby ensuring its safe deployment.

\smallskip
\noindent\fbox{%
    \parbox{\linewidth}{%
In conclusion, GPT-4 serves as a valuable assistant in alleviating the arduous task of constructing extensive safety cases. While it requires human oversight, its contribution in terms of efficiency and reduction of manual effort is noteworthy. GPT-4, in this context, serves as a tool that complements human expertise, facilitating the creation of comprehensive safety cases in a more time-efficient manner.
}
}
\smallskip

\subsection{Can ChatGPT replace the safety case modeler?}

Our preliminary results suggest current state of automation in safety case generation with GPT-4 may come with certain limitations and challenges that need to be acknowledged. Safety-critical systems are characterized by their potential for catastrophic consequences if they fail, making the reliability and accuracy of safety cases paramount. As a result, the idea of fully automating safety case generation, without any human involvement, may not be feasible or advisable at this point. One key concern is that GPT-4 may produce safety cases that, at best, are approximately 80-90 percent structurally and semantically correct. This level of correctness may not sufficiently meet the stringent standards required for safety-critical systems, where precision and accuracy are critical. For this reason, it remains essential --at least for now -- to involve safety case modelers in the safety case generation process. These experts bring a wealth of domain knowledge and years of expertise to the table. They possess critical thinking skills that are invaluable in the argumentation process. By incorporating their input, we can ensure that the generated safety cases not only meet all the structural and semantic correctness criteria but also sufficiently embed domain-specific knowledge.

\smallskip
\noindent\fbox{%
    \parbox{\linewidth}{%

In conclusion, the involvement of safety case modelers in the loop is still needed to bridge the gap between automation and the high standards of safety-critical systems. While automation can expedite the process, human oversight and expertise remain useful in ensuring the highest levels of safety assurance. Still, more experiments on additional safety cases may be needed to generalize our findings.

}
}
\smallskip

\section{Threats to validity} \label{sec7}
\subsection{Internal validity}
The safety cases employed in our experiments were of a relatively modest size. To enhance the robustness and comprehensiveness of our findings, we will conduct in future work additional experiments on larger and more complex safety cases. By doing so, we will be able to further demonstrate the scalability and effectiveness of our approach on a broader spectrum of real-world scenarios.

\subsection{Construction validity}
The contextual information incorporated into the prompts for the experiments conducted in response to RQ2 was derived exclusively from the details provided in the corresponding references (\cite{b27,b28}) about the systems for which the ground-truth safety cases were created. Hence, our knowledge of these  systems is confined to what has been explicitly presented within those references. Any additional or nuanced information pertaining to these systems and that might exist beyond the scope of those papers is not accessible to us. Therefore, it is essential to acknowledge that the safety cases generated by GPT-4 are solely based on the information available within the referenced papers. There is a possibility that critical or contextually significant details about the system may have been omitted or not explicitly addressed in those references.

\subsection{Conclusion validity}
The version of GPT-4 we utilized is known to be non-deterministic, which means that running the model multiple times can yield slightly different results. To ensure the reliability and consistency of our experiments in the future, we will explore the use of a deterministic version of GPT. This deterministic variant may provide stable and predictable outcomes, eliminating the variability associated with non-deterministic models.

\section{Conclusion} \label{sec8}
In this paper, we have explored the possibility of using generative AI to automatically create safety cases. Our investigation has yielded several key findings: (a) GPT-4 demonstrates a commendable level of proficiency in understanding and working with the goal structuring notation (GSN). More specifically, in our assessment, GPT-4 successfully passed with an impressive grade of A. (b) Furthermore, our experiments reveal that GPT-4 possesses the capability to generate safety cases that exhibit a moderate level of accuracy and reasonability. Additionally, it excels in generating safety arguments that bear a striking semantic similarity to the ground-truth safety cases.



Furthermore, considering the manual evaluation process we conducted for assessing the results obtained from GPT-4, which can be quite labor-intensive, we are planning a more streamlined approach for future research. Our future plans involve consolidating the structural and semantic rules that we extracted from the GSN standard. Subsequently, we aim to leverage these rules for the automated assessment of the structural and semantic correctness of safety cases generated by ChatGPT.

Finally, in our future research endeavors, we intend to conduct more experiments on larger safety cases and at exploring the potential synergy between generative AI and other complementary approaches, such as semantic analysis, as Sivakumar et al. suggested \cite{b33}. This synergistic approach has the potential to produce safety cases that exhibit a higher degree of semantic correctness.

\bibliographystyle{ACM-Reference-Format}
\bibliography{main}

\end{document}